\documentclass[useAMS,a4paper,fleqn,usenatbib]{mnras}

\usepackage[T1]{fontenc}
\usepackage{ae,aecompl}
\usepackage[usenames,dvipsnames,svgnames,table]{xcolor}

\usepackage{graphicx}	
\usepackage{amsmath}	
\usepackage{amssymb}	

\usepackage{color}

\title[One-zone model for the Crab nebula broadband spectrum]{Particle acceleration model for the broadband baseline spectrum of the Crab nebula}

\author[F. Fraschetti \& M. Pohl]{
F. Fraschetti,$^{1}$\thanks{E-mail: ffrasche@lpl.arizona.edu}
M. Pohl,$^{2,3}$
\\
$^{1}$Depts. of Planetary Sciences and Astronomy, University of Arizona, Tucson, AZ, 85721, USA\\
$^{2}$DESY, 15738 Zeuthen, Germany\\
$^{3}$Institute of Physics and Astronomy, University of Potsdam, 14476 Potsdam, Germany
}

\date{Accepted XXX. Received YYY; in original form ZZZ}

\pubyear{2017}

\begin{document}
\label{firstpage}
\pagerange{\pageref{firstpage}--\pageref{lastpage}}
\maketitle

\begin{abstract}
We develop a simple one-zone model of the steady-state Crab nebula spectrum encompassing both the radio/soft $X$-ray and the GeV/multi-TeV observations. By solving the transport equation for GeV-TeV electrons injected at the wind termination shock as a log-parabola momentum distribution and evolved via energy losses, we determine analytically the resulting differential energy spectrum of photons. We find an impressive agreement with the observed spectrum of synchrotron emission, and the synchrotron self-Compton component reproduces the previously unexplained broad $200$-GeV peak that matches the Fermi/LAT data beyond $1$ GeV with the MAGIC data. We determine the parameters of the single log-parabola electron injection distribution, in contrast with multiple broken power-law electron spectra proposed in the literature. The resulting photon differential spectrum provides a natural interpretation of the deviation from power-law customarily fit with empirical multiple broken power-laws. Our model can be applied to the radio-to-multi-TeV spectrum of a variety of astrophysical outflows, including pulsar wind nebulae and supernova remnants, as well as to interplanetary shocks.
\end{abstract}




\section{Introduction}

Current theoretical models of the overall spectral energy distribution (SED) of the Crab Nebula fail in reproducing both the peak in the very-high energy (hereafter VHE) range and the radio/$X$-ray data with a single and physically justified population of accelerated particles. The photon differential spectrum from ground-based observatories, i.e., HEGRA \cite{Aharonian.etal:04}, MAGIC \citep{Aleksic.etal:15}, HESS II \citep{Holler.etal:15} and VERITAS \citep{Meagher.etal:15} is consistent with a log-parabola distribution in the range $\sim 50$ GeV up to $100$ TeV. However, the joint photon differential spectrum matching the VHE observations of MAGIC ($0.05-30$ TeV) with the Fermi/LAT spectrum in the range $1-200$ GeV exhibits a broad and flat peak that is fit \citep{Aleksic.etal:15} by a modified log-parabola with an additional {\it ad hoc} free-parameter. Such a VHE spectrum cannot originate from an electron-positron spectrum evolved from the standard power-law distribution of shock-accelerated energetic particles; however, no attempt has been reported so far to reconstruct the electron-positron source spectrum.

The VHE emission of the Crab nebula is usually modelled with Inverse Compton (hereafter IC) radiation from a non-thermal population of electron-positron pairs via scattering of three possible photon targets: synchrotron radiation emitted by the same electron-positron population within the nebula (Synchrotron self-Compton) in the spectral range from the optical to $X$ rays, thermal far-infrared radiation (typically associated with dust) or cosmic microwave background (CMB) radiation. A variety of radiation processes proposed to explain the broadband spectrum of the Crab pulsar wind nebula (PWN) seems to fail in matching the VHE emission with the emission at smaller wavelengths from radio to $X$-rays, as discussed in \cite{Aleksic.etal:15}: the MHD spherical flow model of the downstream pulsar wind structure \citep{Kennel.Coroniti:84,Meyer.etal:10} fits well only the synchrotron part, not the VHE part of the spectrum; a simplified model with uniform magnetic field throughout the nebula and two electron populations (relic of the pulsar wind, that cooled to produce synchrotron radio, and wind) fails to reproduce the $200$ GeV peak; a time-dependent model solving the diffusion-loss equation that includes synchrotron, IC and bremsstrahlung losses fails to reproduce the size of the nebula inferred at smaller wavelengths. Thus, the determination of the energetic particles distribution leading to the observed $\gamma$-ray spectrum and compatible with the low-energy spectrum (radio and soft $X$-rays) is an open problem. 

Within the PWN standard scenario, the magnetized wind from the pulsar advects electron-positron pairs presumably \citep{Arons.Tavani:94} accelerated to a power-law distribution by the wind termination shock (hereafter TS) via a diffusive shock-acceleration {\it \'a} la Fermi. An alternative scenario based on shock-driven reconnection in the striped wind was proposed by \cite{Petri.Lyubarski:07}. In this paper we adopt the former scenario and show that a physically motivated energetic particles distribution can explain the broadband observations. 

Energetic particle spectra measured {\it in-situ} at interplanetary shocks exhibit in several cases deviation from a single power-law. A multispacecraft analysis of the energetic ions spectra \citep{Mewaldt.etal:12} reported that all the $16$ Ground Level Events during solar cycle $23$ can be fit by broken power laws or by the Band function \citep{Band.et.al:93} in the range $0.1$ to $500-700$ MeV with break energies ranging from $\sim 2$ to $\sim 46$ MeV. This result suggests that no-power-law spectra over $2$ or $3$ energy decades are perhaps a commonplace for shocks.

In this paper we investigate the microscopic origin of the broadband photon spectrum of the Crab nebula. 
We show that a single population of electrons evolved via energy losses from a log-parabola injection spectrum leads to: 1) quasi-log-parabola photon differential spectrum produced via synchrotron self-Compton scattering and IC scattering over CMB in agreement with the overall $\gamma-$ray observations between $1$ GeV and $100$ TeV; 2) low-energy synchrotron spectrum in the range $(10^{-5}- 10^4)$ eV in agreement with radio/$X$-rays observations (also reported in \citet{Aleksic.etal:15}). The injection of the energetic electrons is found to decrease at a very slow rate over the entire lifetime of the nebula ($\sim 10^3$ years). We provide a probabilistic microscopic interpretation of the underlying process of particle acceleration at the Crab nebula TS.

This paper is organised as follows. In Sect. \ref{outline} we outline the basic assumptions of the microscopic model for the energetic particles. In Sect. \ref{model} the model for the synchrotron radiation and the IC scattering emission is described in detail. In Sect. \ref{spectrum} we compare the predicted photon spectrum with the broadband observations, contrast the log-parabola electron injection distribution with asymptotic power-laws. In Sect. \ref{discussion} we derive the relation between the parameters of the electrons distribution and the observed photon spectrum and estimate the diffusion-advection scale upstream of the TS. Sect \ref{conclusions} summarises our findings.

\section{Outline of the model}\label{outline}

For a population of energetic particles at an infinitely planar collisionless shock wave, the process of acceleration is self-sustained if individual particles, advected downstream after shock-crossing, can contrast the flow advection and return to the shock to undergo further acceleration. The reservoir of far downstream particles can be phenomenologically treated by introducing the probability of return, $P(v)$, i.e., the probability that an individual particle with speed $v$ in the frame of the downstream fluid can return to the shock\footnote{By definition, $P (v)$ is the fraction of the particle flux that crosses the shock from upstream to downstream and return to the shock over the total flux of particles crossing from upstream to downstream.}. 
In contrast, for the upstream fluid, all particles penetrating into the upstream are overtaken by the shock, unless turbulence is unable to confine them and they escape to a far-away observer; the escape can be mathematically modelled by requiring a free escape boundary whose location upstream is generally treated as a free parameter.

Numerical simulations have extensively used $P (v)$ by implementing the scattering off the magnetic fluctuations with a Monte-Carlo method \citep{Ellison.etal:96} or by synthesizing the magnetic turbulence with a prescribed power spectrum \citep{Giacalone:05,Fraschetti.Giacalone:15}. For flows crossed by a non-relativistic shock wave (e.g., interplanetary or supernova remnant shock), if $U_2$ is the speed of the downstream plasma in the shock rest frame, such probability reads $P (v) =[(v-U_2)/(v+U_2)]^2$, increasing as a function of $v$ and independent of energy for $v \rightarrow c$ \citep[e.g.]{Jones.Ellison:97}. 
The definition of $P (v)$ requires the isotropy of the pitch-angle distribution in the local plasma frame, or in other terms that $v$ is so much greater than the flow speed $U$ that second order terms in $U/v$ can be disregarded. We note that, if the condition $v \gg U$ is not satisfied, i.e., at supra-thermal energy, particles are not fast enough to isotropize their pitch-angle distribution in the downstream plasma frame close to the shock. \footnote{An example is the emergence of spikes in the intensity profiles of supra-thermal particles measured across interplanetary shocks \citep{Lario.etal:03}, explained as result of ion reflection by the magnetic barrier at quasi-perpendicular shocks \citep{Gieseler.etal:99} or pile-up of particles streaming along turbulent field lines crossing the shock surface multiple times \citep[and references therein]{Fraschetti.Giacalone:15}. As a result, the supra-thermal particle spectrum downstream close to the shock is harder than the one-dimensional prediction of the diffusive shock acceleration model.} The same expression of $P (v)$ holds for ultra-relativistic shocks \citep{Peacock:81} and was extended to any relativistic shock speed \citep{Summerlin.Baring:12}, provided that the particle distribution is isotropic in the downstream plasma frame.

The momentum gain relates differently to the flow speed between ultra-relativistic flows and non-relativistic flows. For an isotropically distributed incoming population of relativistic electrons crossing a quasi-perpendicular ultra-relativistic shock, such as a pulsar wind TS at large-scale, the momentum gain upon the first cycle of crossing and re-crossing the shock is $p_f / p_i \simeq \Gamma^2/2$, where the initial ($p_i$) and final ($p_f$) momentum are calculated in the upstream fluid frame and $\Gamma$ is the Lorentz factor of the upstream fluid in the shock frame. For the subsequent shock-crossings, $p_f / p_i $ is limited by geometrical constraints to smaller values \citep{Gallant.Achterberg:99}. For a non-relativistic fluid, the momentum gain upon first cycle of shock crossing is smaller than for ultra-relativistic flows. Upon any cycle of shock crossing, including the first one, $\Delta p = p_f - p_i $ can be written, by averaging over an isotropic distribution, as $\langle \Delta p / p \rangle  = 4/3 \, \Delta U/v$, where $\Delta U$ is the change of flow speed across the shock, and $v$ is the particle speed ($v$ and $p$ are computed in the local downstream plasma frame). This expression for $\Delta p / p$ is valid only if higher-order terms in $U/v$ are neglected.

Several effects can hamper the energy gain of charged particles interacting with shock waves.  
In the case of a large-scale quasi-perpendicular magnetic obliquity, such as the Crab TS, the acceleration proceeds as long as the transport in the direction perpendicular to the average field due either to field line meandering \citep{Jokipii:66} or to the motion across the local perturbed field lines \citep{Fraschetti.Jokipii:11} allows particles to re-cross the shock. Ultra-relativistic shocks, such as presumably a pulsar wind TS, have predominantly perpendicular obliquity leading to the high compression of the magnetic component perpendicular to the average direction of the shock motion; moreover, the particle speed and the shock speed are very close, in contrast with non-relativistic shocks, so that the isotropy in the downstream medium can be attained only far from the shock. Therefore, particles are more likely to be lost in the downstream of a shock at ultra-relativistic than at non-relativistic flow speeds. 

In this paper we propose that the Crab wind TS accelerates to the multi-TeV range a single population of electrons that produces the observed baseline differential spectrum between the radio and the VHE band. We consider here the realistic case that the shock cannot be assumed infinitely planar at the typical spatial scale of particle motion (i.e., gyroradius $r_g$) of the highest-energy particles ($\sim$ TeV) as a result of inherent geometrical effects limiting the particle acceleration, for instance the corrugation at various scales of the shock surface or its large-scale curvature; the gyroscale is usually neglected as typically much smaller than the other scales. As the particle speed grows, such geometrical effects modify the probability that a particle remains in proximity of the shock introducing a mild energy-dependence, that accounts for the highest-energy particles with $r_g$ relatively large to leak out of the system, thereby leading to a softening of the photon spectrum.

Throughout the paper we shall assume for simplicity that the energy gain per cycle is a constant fraction of the particle energy, i.e. the energy gain is linear in energy. 
Then, we consider the probability ${\cal P} (\gamma)$, where $\gamma$ is the particle Lorentz factor in the downstream plasma frame where the observed radiation is emitted, as the probability of remaining in the acceleration region, both upstream or downstream, that incorporates the probability of return to the shock from downstream, the geometrical factors limiting the acceleration listed above and the particle escape upstream.   
For such VHE electrons we use the microscopic {\it ansatz} that ${\cal P} (\gamma)$ depends on $\gamma$ as ${\cal P} (\gamma)= g/\gamma^q$, where $g$ is a normalisation factor and $0<q <1$ a parameter. Such an expression does not assume isotropy in the local plasma frame. The anisotropy expected at ultra-relativistic flows in the downstream local frame behind and very close to the shock, as compared to the gyroscale of $\simeq 1$ TeV electrons downstream, does not contradict the {\it ansatz} ${\cal P}(\gamma) = g/\gamma^q$.  
Test-particle simulations show that isotropy in the local downstream plasma frame is recovered \citep{Summerlin.Baring:12}.

 It has been shown that such {\it ansatz} leads to a log-parabola spectrum of energetic electrons \citep{Massaro.etal:04} that produce a quasi-log-parabola synchrotron spectra, modulo a logarithmic factor, in good agreement with the steady $X$-ray spectra observed from blazars. Moreover, a significant fraction of photon spectra of extended sources listed in the 3FGL catalog are better represented by a log-parabola functional form than with a single power-law \citep{Fermicoll.15}.

We define $Q(> \gamma)$ as the number density of energetic electrons accelerated to a total energy higher than $m_e \gamma c^2$, where $m_e$ is the electron mass, and $R=\gamma_i/\gamma_{i-1}$ as the ensemble-averaged increment in energy at the $i^{th}$ shock-crossing, given by the ratio of the $i^{th}$, $\gamma_i$, to the previous, $\gamma_{i-1}$, particle Lorentz factor. The form ${\cal P} (\gamma)= g/\gamma^q$ yields\footnote{If the probability of return ${P} (v)$ is independent of the particle energy, the power-law differential spectrum $\sim N_0 \gamma^{-p}$ is re-obtained, where the index $p = -{\rm ln} P(v)/{\rm ln} R$ can be re-written in terms of the shock density compression ratio in the test-particle regime. If the backreaction of energetic particles grows as to modify the upstream turbulence, the spectrum will be concave, instead of a simple power-law, as particles with different energy  experience different shock compression.} \citep{Massaro.etal:04}
\begin{equation}
Q(> \gamma) = N_0 (\gamma/ \gamma_0)^{-[s-1+r{\rm log}(\gamma/ \gamma_0)]}
\label{N}
\end{equation}
where $N_0$ is a normalization constant, $\gamma_0$ is the electron injection factor and the spectral parameters $s$ and $r$ are related to the microscopic statistical parameters $g, q, R$ as follows: $s = -{\rm log}(g/\gamma_0^q)/{\rm log}(R) - (q-2)/2$, $r=q/2 {\rm log}(R)$. It is noteworthy clarifying that $\gamma_0$ is the injection threshold into the process of acceleration limited by geometrical factors in the VHE range, and not the injection energy from the supra-thermal pool of particles that undergo diffusive shock acceleration.

\section{Theoretical broadband flux from log-parabola electron distribution}\label{model}

We assume that the non-thermal population of energetic electrons/positrons pairs is produced with a log-parabola injection spectrum in Eq. \ref{N}, accounting for a slowly decreasing rate, at the pulsar wind TS up to energies $\sim 100$ TeV, evolves under the influence of energy losses, and feeds the observed baseline photon emission from the radio up to the multi-TeV range. We calculate analytically in this section the total synchrotron and IC fluxes including energy losses during the entire nebula life-time. All quantities are calculated in the reference frame of the fluid downstream of the TS, i.e., the nebula, where synchrotron photons are emitted by the multi-TeV electrons and, in part, upscattered to VHE range.

\subsection{Time dependence of the electron spectrum}\label{tdep}

In this sub-section we provide an analytic expression for the spectrum of electrons that results from injection at the TS with the spectrum given by Eq.\ref{N}. The GeV-TeV electrons are assumed to be injected in a spatially uniform fashion at the source, i.e., Crab PWN TS, so that the differential electron number spectrum $\bar N (\gamma, t)$ is independent of location. Therefore, $\bar N (\gamma, t)$ satisfies the simplified transport equation
\begin{equation}
\frac{\partial \bar N (\gamma, t)}{\partial t}- \frac{\partial}{\partial \gamma} (b\gamma^2\, \bar N (\gamma, t)) = Q(\gamma, t),
\label{transp_eq}
\end{equation}
with an energy loss rate $-b\,\gamma^2$ and a differential injection rate of the electrons $Q(\gamma, t)$ at time $t$. 

From Eq. \ref{N}, the differential electron production rate $Q(\gamma) = d Q (> \gamma) / d \gamma$ is given by
\begin{equation}
Q(\gamma) = \frac{N_0}{\gamma_0}\,\left(\frac{\gamma}{\gamma_0}\right)^{-s-r\log\left(\frac{\gamma}{\gamma_0}\right)}
\, {\left|s-1+2r\log\left(\frac{\gamma}{\gamma_0}\right)\right|} ,
\label{Q_gamma}
\end{equation}
that is a log-parabola multiplied by a pre-factor logarithmically dependent on $\gamma/ \gamma_0$. The injection term in Eq. \ref{transp_eq} is taken to have the form 
\begin{equation}
Q(\gamma,t)=Q(\gamma)\,(1+a)\,\frac{t^a}{\tau_0^{1+a}}\,\Theta\left(\tau_0-t\right)\,\Theta\left(t\right)\ ,
\label{Q_a}
\end{equation}
where $\tau_0 \simeq 10^3$ years denotes the current age of the remnant, and $a < 0$ is a free parameter modulating the electron injection rate; $\Theta(x)$ indicates the Heaviside function.

Defining the synchrotron-loss time scale as $\tau_\mathrm{syn}=1/b\gamma_0$, we find for the differential electron number spectrum at the present time, $\bar N(\gamma, \tau_0) \equiv N(\gamma)$
\begin{align}
N(\gamma)&=\frac{1}{b\,\gamma^2}\int_\gamma^{\gamma_\mathrm{max}} d\gamma'\ \int_0^{\tau_0} dt'\ Q(t',\gamma')
\,\delta\left(\tau_0-t'+\frac{\gamma-\gamma'}{b\,\gamma\,\gamma'}\right) \nonumber\\
&=\frac{1+a}{b\,\gamma^2\,\tau_0^{1+a}}\int_\gamma^{\gamma_\mathrm{max}} d\gamma'\ Q(\gamma')\,\left[\tau_0 +\frac{\gamma-\gamma'}{b\,\gamma\,\gamma'}\right]^a \nonumber\\
&=(1+a)\,N_0\frac{\tau_\mathrm{syn}}{\tau_0}
\frac{1}{\gamma}\int_1^{x_\mathrm{max}} dx\ \left[1 +\frac{(1-x)\,\gamma_0\,\tau_\mathrm{syn}}{\gamma\,x\,\tau_0}\right]^a\, \nonumber\\ 
& \quad \times \left(\frac{\gamma\,x}{\gamma_0}\right)^{-s-r\log\left(\frac{\gamma\,x}{\gamma_0}\right)}   
{\left|s-1+2r\log\left(\frac{\gamma x}{\gamma_0}\right)\right|}\, ,
\label{espec}
\end{align}
where $\delta (\cdot)$ in the first line of Eq. \ref{espec} is the Dirac-delta function and we have used $x=\gamma'/\gamma$. The upper limit of integration is
\begin{equation}
\gamma_\mathrm{max}=\begin{cases}\frac{\gamma}{1-b\,\gamma\,\tau_0}&\mbox{if } \gamma<\frac{1}{b\,\tau_0} \\
\infty &\mbox{otherwise}\end{cases}
\end{equation}
or
\begin{equation}
x_\mathrm{max}=\begin{cases}\frac{1}{1-b\,\gamma\,\tau_0}&\mbox{if } \gamma<\frac{1}{b\,\tau_0} \\
\infty &\mbox{otherwise}\end{cases}\ .
\end{equation}

\subsection{Synchrotron radiation}\label{sync_tdep}

The synchrotron power emitted by a single electron, averaged over an isotropic electron distribution, is given in the local plasma frame by 
$P_\mathrm{syn} = (\sigma_T c / 6 \pi) \gamma^2 B^2$, where $\sigma_T = ({8\pi/3})({e^4/m_e^2 c^4})$ is the Thomson cross-section, and $B$ is the external magnetic field assumed to be uniformly embedded in the nebula. The total synchrotron flux at Earth from a source at distance $d$, namely $\nu F_{\nu}^\mathrm{syn}$, is found by folding $P_\mathrm{syn}$ with the differential electron spectrum in Eq. \ref{espec}: $\nu F_{\nu}^\mathrm{syn} = {1 \over 4\pi d^2} \int d \gamma P_\mathrm{syn} N (\gamma)$. 
We use the monochromatic approximation, i.e., the electron power is concentrated around the characteristic synchrotron frequency $\nu_{s} = 0.29 (3e\gamma^2 B)/(4 \pi m_e c) = \gamma^2 \nu_s^0$. Thus, the total flux can be recast as $\nu F_{\nu}^\mathrm{syn} \simeq {1 \over 4\pi d^2} \int d \gamma P_\mathrm{syn} N (\gamma)|_{\nu = \nu_s}$  which leads to
\begin{equation}
\nu F_{\nu}^\mathrm{syn} (\epsilon) = \frac{\sigma_T\,c\,B^2}{12\pi\,4\pi\,d^2}\
\left(\frac{\epsilon}{\epsilon_0}\right)^{3/2}\,
N\left(\sqrt{\frac{\epsilon}{\epsilon_0}}\right)\ ,
\end{equation} 
where we have used ${\epsilon_0} = h \nu_s^0$ with the Planck constant, $h$. Inserting the expression for $N (\gamma)$ in Eq. \ref{espec} we find for the total synchrotron flux  
\begin{align}
\nu F_{\nu}^\mathrm{syn} (\epsilon)= &\frac{\sigma_T\,c\,B^2}{12\pi\,4\pi\,d^2}\,\frac{(1+a)\,N_0\,\tau_\mathrm{syn}}{\tau_0}\,\gamma_0^2\,
\left(\frac{\epsilon}{\epsilon_s}\right)\,\nonumber\\
&\times\int_1^{x_\mathrm{max}} dx\ \left[1 +\sqrt{\frac{\epsilon_s}{\epsilon}}\frac{\tau_\mathrm{syn}}{\tau_0}\frac{1-x}{x}\right]^a\, \nonumber\\
& \times \left(x\,\sqrt{\frac{\epsilon}{\epsilon_s}}\right)^{-s-r 2\log\left(x\sqrt{\frac{\epsilon}{\epsilon_s}}\right)} 
 \left |s-1+2r{\rm log}\left(x\,\sqrt{\frac{\epsilon}{\epsilon_s}}\right)\right| \ ,
 \label{f_syn_tdep}
\end{align}
where we have used $\epsilon_s=\gamma_0^2\,\epsilon_0$. To retain consistency in notation we write
\begin{equation}
x_\mathrm{max}=\begin{cases}\left(1-\sqrt{\frac{\epsilon}{\epsilon_s}}\frac{\tau_0}{\tau_\mathrm{syn}}\right)^{-1}&\mbox{if } \epsilon<\epsilon_s\left(\frac{\tau_\mathrm{syn}}{\tau_0}\right)^2 \\
\infty &\mbox{otherwise}\end{cases}\ .
\end{equation}

Equation \ref{f_syn_tdep} contains a log-parabola distribution in $x\,\sqrt{{\epsilon}/{\epsilon_s}}$, first factor in the third line of Eq. \ref{f_syn_tdep}, inside the $x$-integrand that takes into account the energy losses due to synchrotron radiation during the entire nebula lifetime. The factor in $|\cdot|$ in the third line of Eq. \ref{f_syn_tdep} is usually neglected in the fit of blazars $X$-ray spectra, as only logarithmically increasing; such factor controls the width of the spectral peaks at $\sim 1$ eV and $\sim 200$ GeV. 
The flux given in Eq. \ref{f_syn_tdep} will be compared in the next section with radio and $X$-ray data for the Crab nebula baseline spectrum.

\subsection{Synchrotron self-Compton and inverse Compton scattering from CMB radiation}\label{IC_tdep}

In this sub-section we analytically calculate the predicted IC flux due to scattering off the synchrotron photons determined in Sect. \ref{sync_tdep} and off CMB radiation within the nebula. The Lorentz-invariant differential density of synchrotron photons inside a spherical source of radius $R$, $n(\epsilon)=dn/(dV \, d\epsilon)$, is for optically thin systems related to the mean intensity, $J_\nu$, and that to the emission coefficient, $j_\nu$, as
\begin{align}
&n_s (\epsilon)=\frac{4\pi}{c\,h\,\epsilon}\,J_\nu (\epsilon) \quad ;\quad J_\nu (\epsilon) = \frac{3\,R}{4}\,j_\nu (\epsilon)\nonumber\\
& \Rightarrow\ n_s(\epsilon)=\frac{3\pi\,R}{c\,h\,\epsilon}\,j_\nu (\epsilon) ,
\end{align}
where $\epsilon$ is the target photon energy in the local plasma frame. The total synchrotron flux at Earth depends on the emission coefficient as
\begin{equation}
\nu F_{\nu}^\mathrm{syn} (\epsilon) = \frac{4\pi}{3}\,\frac{R^3}{d^2}\nu \,j_\nu (\epsilon)
\end{equation}
Thus, the differential photon density of synchrotron photons is
\begin{align}
n_s(\epsilon) 
&=\frac{9}{4\,c\,\epsilon^2}\,\frac{d^2}{R^2}\,\nu F_{\nu}^\mathrm{syn} (\epsilon)  \nonumber\\
&=\frac{3\,\sigma_T\,B^2}{64\pi^2\,R^2}\,\frac{(1+a)\,N_0\,\tau_\mathrm{syn}}{\tau_0\,\epsilon_s^2}\,\gamma_0^2\,
\left(\frac{\epsilon}{\epsilon_s}\right)^{-1}\,\nonumber\\
&\quad\times\int_1^{x_\mathrm{max}} dx\ \left[1 +\sqrt{\frac{\epsilon_s}{\epsilon}}\frac{\tau_\mathrm{syn}}{\tau_0}\frac{1-x}{x}\right]^a\,
\left(x\,\sqrt{\frac{\epsilon}{\epsilon_s}}\right)^{-s-r\log\left(x\sqrt{\frac{\epsilon}{\epsilon_s}}\right)} \nonumber\\
&\quad \times \left |s-1+2r{\rm log}\left(x\,\sqrt{\frac{\epsilon}{\epsilon_s}}\right)\right|
\ .
\end{align}
where in the last step we have used Eq. \ref{f_syn_tdep}. The total differential photon density includes $n_s (\epsilon)$ and the differential CMB photon density $n_\mathrm{CMB}$: $n_{\rm tot} (\epsilon) = n_s (\epsilon) + n_\mathrm{CMB}$.

For an ambient photon population isotropic in the local plasma frame with a differential density $n(\epsilon)$, the spectrum of the scattered photons with initial energy between $\epsilon $ and $\epsilon + d \epsilon$ and final energy $\epsilon_1$ per single electron is \citep{Blumenthal.Gould:70} 
\begin{equation}
{d N_{\gamma, e} \over dt d\epsilon_1}  = {3 \sigma_T c \over 4\gamma^2}  {n_{\rm tot}(\epsilon)\over \epsilon} d\epsilon G(q, \bar \Gamma),
\label{Ngamma}
\end{equation}
with
\begin{equation}
G(q, \bar \Gamma) = 2q{\rm ln} q + (1+2q)(1-q) + \frac{\bar \Gamma^2 q^2 (1-q)}{2(1+\bar \Gamma q)}
\label{G}
\end{equation}
and 
\begin{equation}
\bar \Gamma = \frac{4\epsilon \gamma}{ m_e c^2}  ,\quad  q = \frac{\epsilon_1}{\bar \Gamma(m_e \gamma c^2 - \epsilon_1)}.
\end{equation}
where Eqs (\ref{Ngamma}, \ref{G}) are derived under the only restriction $\gamma \gg 1$. The calculation of the IC photon number emitted per unit time by a single electron in the local plasma frame over the target photon distribution $n(\epsilon)$, namely $\Pi_\mathrm{IC} (\gamma, \epsilon_1)$, leads to 
\begin{eqnarray}
\Pi_\mathrm{IC} (\gamma, \epsilon_1) & =& \int_0^\infty d\epsilon {d N_{\gamma, e} \over dt d\epsilon_1 d\epsilon}  \epsilon_1  \nonumber\\
&=& {3 \sigma_T c \over 4\gamma^2} \epsilon_1 \int_0^\infty d\epsilon \, \frac{n_{\rm tot}(\epsilon)}{\epsilon} G(q, \bar \Gamma) .
\label{P_IC}
\end{eqnarray}
We note that in the $\epsilon$-integral of Eq.\ref{P_IC} we calculate the contribution of target synchrotron photons at all frequencies with no monochromatic approximation. Since the synchrotron photon energies extend over a wide range ($10^{-5}$ eV up to $100$ keV), Eq. \ref{P_IC} entails scattering at all regimes between Thomson ($\bar \Gamma \ll 1$) and Klein-Nishima ($\bar \Gamma \gg 1$). 
The total IC photon number emitted per unit time by the electron population is found by folding Eq. \ref{P_IC} with differential distribution $N(\gamma)$ in Eq.\ref{espec} $\Pi_\mathrm{IC} (\epsilon_1) = \int {d\gamma} \Pi_\mathrm{IC} (\gamma, \epsilon_1) N (\gamma) $. 

\begin{figure*}
	\includegraphics[width=0.8\textwidth]{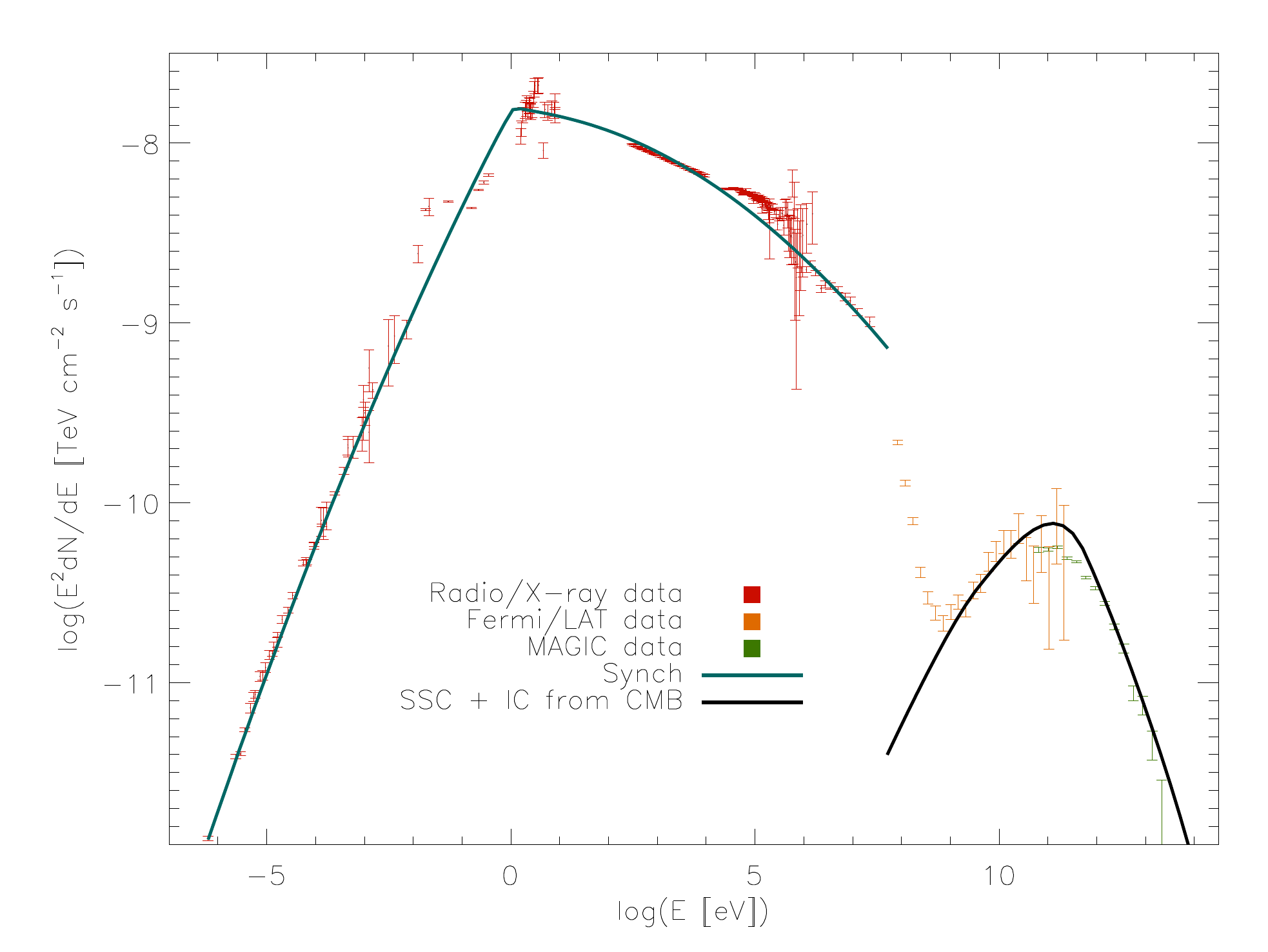}
\caption{Theoretical differential synchrotron photon spectrum (cyan) in Eq. \ref{f_syn_tdep} compared with observed radio and soft $X$-rays spectra from ISO-SCUBA, Spitzer, XMM-Newton, SPI, IBIS/ISGRI (same set as used in \citet{Aleksic.etal:15}) and theoretical IC spectrum (black) in Eq. \ref{F_IC_TOT_tdep} compared with the joint Fermi/LAT and MAGIC \citep{Aleksic.etal:15} spectra.}
\label{F_TOT_tdep}
\end{figure*}

The observed IC energy flux is related to $\Pi_\mathrm{IC}(\epsilon_1)$ by 
\begin{align}
\nu F_{\mathrm{IC},\nu}^{\rm tot} (\epsilon_1) &=\frac{\epsilon_1}{4\pi\,d^2}\,\Pi_\mathrm{IC}(\epsilon_1)
 \nonumber \\
&=\epsilon_1^2\,\frac{3\,c\,\sigma_T}{16\pi\,d^2}\,
\int d\epsilon\ \frac{n_{\rm tot}(\epsilon)}{\epsilon}\,
\int d\gamma\ \frac{N(\gamma)}{\gamma^2}\,G(q,\bar \Gamma) 
\label{F_IC_TOT_tdep}
\end{align}
where the term due to synchrotron self-Compton can be re-written as 
\begin{align}
\nu F_{\mathrm{IC},\nu}^{\rm ssc} (\epsilon_1) =\epsilon_1^2\,\frac{27\,\sigma_T}{64\pi\,R^2}\,
\int d\epsilon\ \frac{\nu F_{\nu}^\mathrm{syn} (\epsilon)}{\epsilon^3}\,
\int d\gamma\ \frac{N(\gamma)}{\gamma^2}\,G(q,\bar \Gamma) \ .
\end{align}
We emphasize that the energy fluxes in the radio/$X$-ray band (Eq. \ref{f_syn_tdep}) and in the VHE band (Eq. \ref{F_IC_TOT_tdep}) use the same differential energetic electron distribution (Eq. \ref{espec}). Those energy fluxes will be compared in the next section with the baseline broadband spectrum of the Crab nebula.

\section{Comparison with broadband observations}\label{spectrum}

\begin{figure}
	\includegraphics[width=0.5\textwidth]{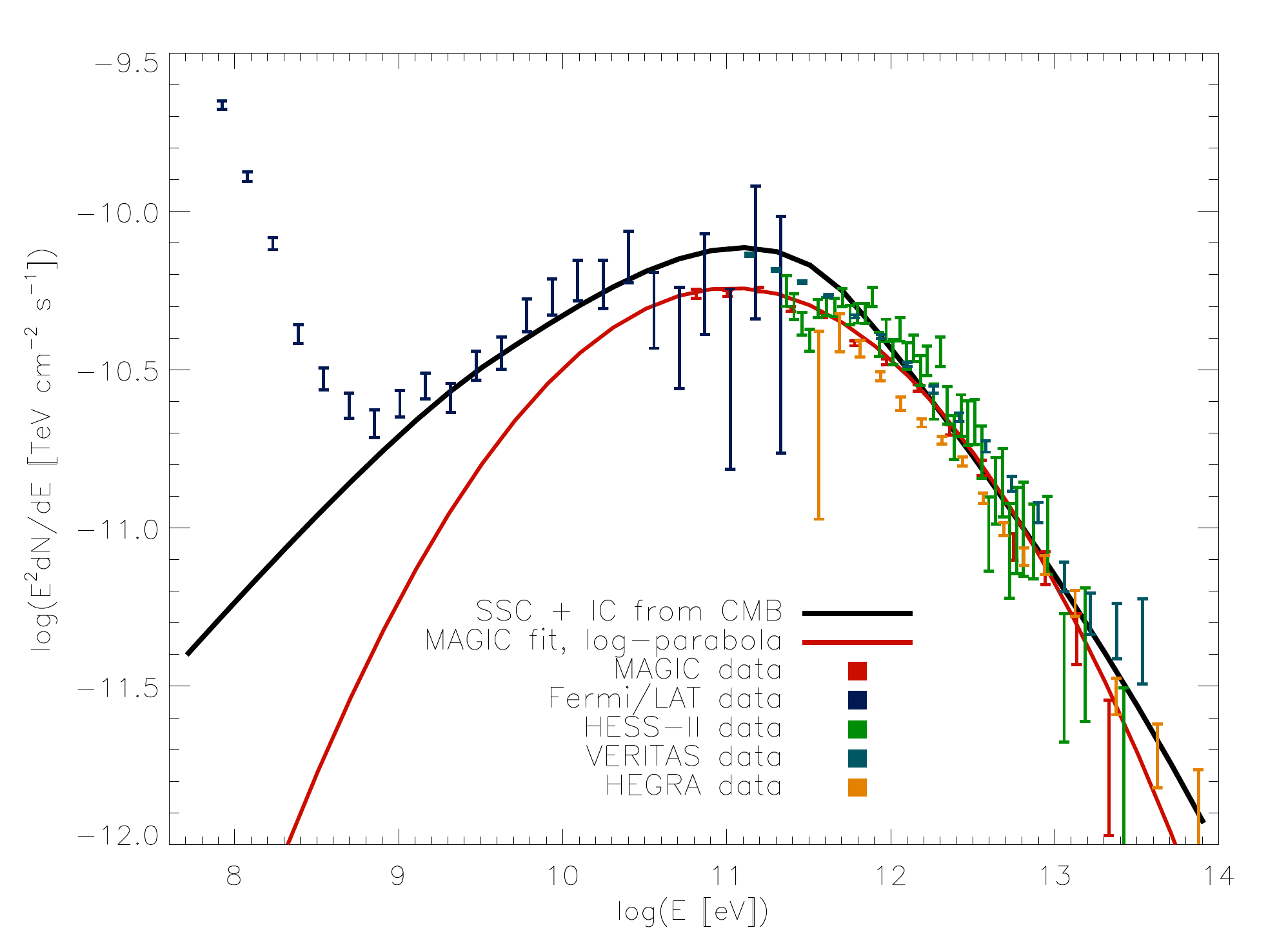}
\caption{Theoretical IC spectrum (black) as in Eq. \ref{F_IC_TOT_tdep} compared with VHE band spectra from various ground-based observatories: HEGRA \citep[$300$ GeV- $100$ TeV]{Aharonian.etal:04}, MAGIC \citep[$50$ GeV - $30$ TeV]{Aleksic.etal:15}, HESS II \citep[ $230$ GeV - $25$ TeV]{Holler.etal:15} and VERITAS \citep[$115$ GeV - $42$ TeV]{Meagher.etal:15}. The parameters of the theoretical spectrum are the same as in Fig.\ref{F_TOT_tdep}. The log-parabola fit of the MAGIC data reported in \citet{Aleksic.etal:15} is represented by the red curve. }
\label{F_VHE}
\end{figure}

\begin{figure}
	\includegraphics[width=0.5\textwidth]{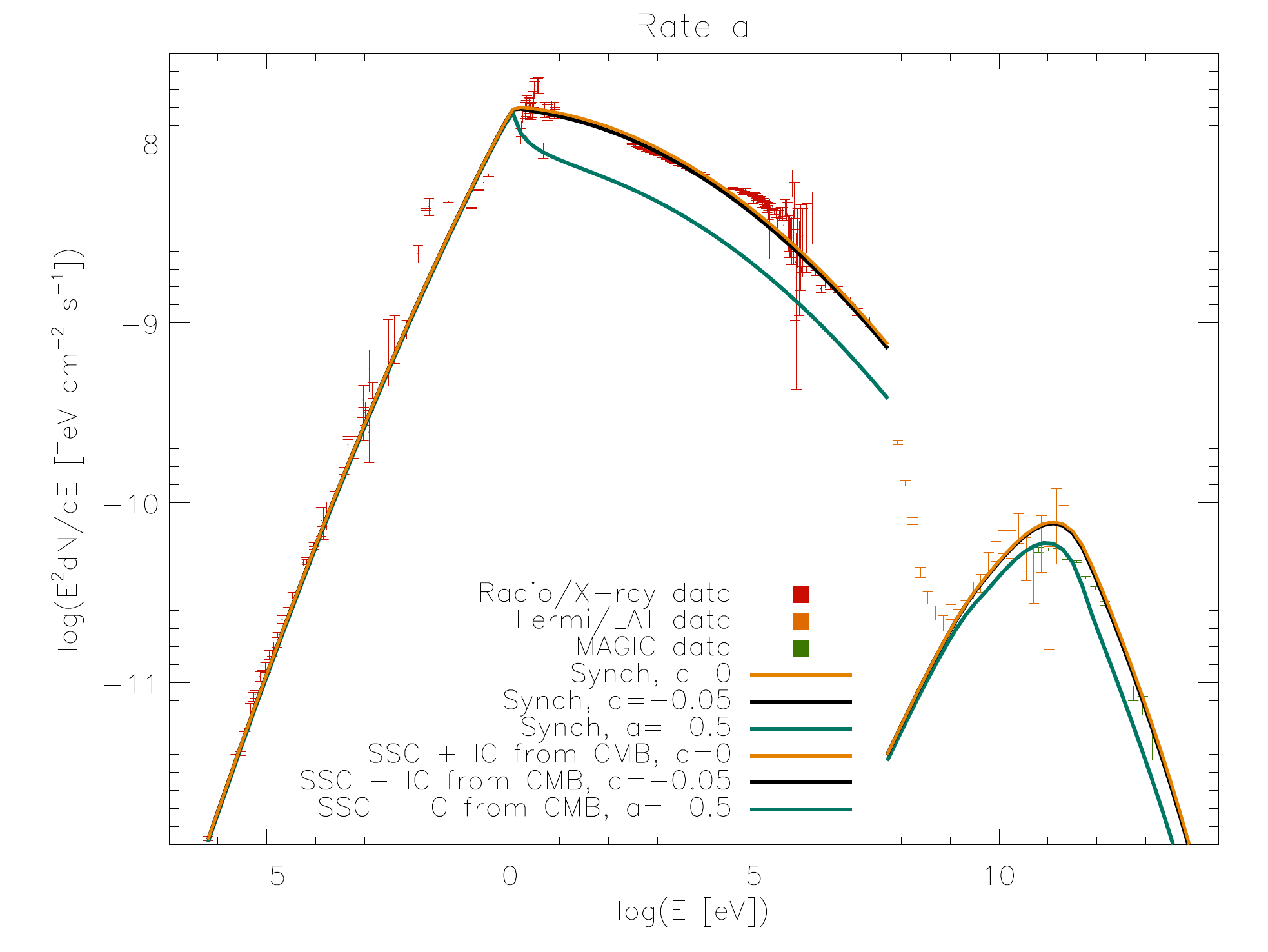}
\caption{Theoretical synchrotron photon spectra (Eq. \ref{f_syn_tdep}) and IC spectra (Eq. \ref{F_IC_TOT_tdep}) for three values of $a$ ($-0.5, -0.05, 0$) are compared with the same data sets as in Fig.\ref{F_TOT_tdep}. For the other parameters the best-fit value in Fig.\ref{F_TOT_tdep} was used.}
\label{F_a_3}
\end{figure}

Figure \ref{F_TOT_tdep} compares the theoretical differential spectrum of synchrotron photons given by Eq. \ref{f_syn_tdep} with the observed radio and soft $X$-ray spectra ($10^{-5}$ eV up to $\sim 10$ MeV), and likewise the theoretical IC radiation as in Eq. \ref{F_IC_TOT_tdep} with the joint Fermi/LAT and MAGIC spectra. The best-fit involves $4$ free parameters for the overall spectrum: $s = 1.75$, $r=0.08$, $N_0 = 2.6 \times 10^{49}$, and $a=-0.05$ (see also Fig. \ref{F_a_3}) for an energy scale $\gamma_0 = 2 \times 10^4$. We have used a magnetic-field strength $B$ constant and uniform throughout the nebula of the order of the typical value, i.e., $B = 140 \, \mu$G (cfr. the best-fit value for the constant $B$-field model in \cite{Aleksic.etal:15}, $B=143 \,\mu$G).  We notice that for this magnetic-field strength the synchrotron losses dominate for $\gamma' \gtrsim 10^2 \gamma_0 = 2 \times 10^6$, leading to $\tau_\mathrm{syn} \lesssim \tau_0$; at smaller electron energy the loss time is larger than the age of the Nebula itself. As a consequence the synchrotron flux peaks at a photon energy $\epsilon =  \gamma'^2  \epsilon_0 = 2.7$ eV and diminishes at higher energy (see Fig. \ref{F_TOT_tdep}). Finally, we have used a size of the nebula $R = 1$ pc and a distance from the observer $d = 1.9$ kpc.

The VHE region of the spectrum is zoomed-in in Fig.\ref{F_VHE} and compared with reconstructed spectra obtained by a number of VHE ground-based observatories. The overlaid log-parabola fit of the MAGIC data only (red in Fig. \ref{F_VHE}) reported in \citep{Aleksic.etal:15} has the form
\begin{equation}
\frac{d {\cal N}}{d\varepsilon dA dt} (\varepsilon) \propto (\varepsilon/ \varepsilon_0^M)^{-\alpha+\beta \cdot{\rm log}(\varepsilon/ \varepsilon_0^M)} 
\label{MAGIC_spectrum}
\end{equation}
with index $\alpha = 2.47  \pm 0.01 $, curvature $\beta = -0.24 \pm 0.01$ and scale $\varepsilon_0^M = 1 $ TeV (note that $\varepsilon_0^M \simeq m_e c^2 \gamma'$). 

Figure \ref{F_VHE} shows that MAGIC's empirical log-parabola in Eq. \ref{MAGIC_spectrum} and our solution in Eq. \ref{F_IC_TOT_tdep} equally well reproduce the MAGIC data down to the IC peak at $\simeq 200$ GeV. However, a pure log-parabola curve fails to explain the lower energy (Fermi/LAT) spectrum down to $\sim 1$ GeV, that \citet{Aleksic.etal:15} proposes to fit with an empirically modified log-parabola relying on an additional fitting parameter.
In contrast, we show that the theoretical physically motivated curve derived here is in satisfactory agreement with the Fermi/LAT rising part of the spectrum, the broad $200$ GeV peak and the VHE spectrum. We note that the value $\tau_0 = 1,000$ years used here is a crude approximation and might be smaller due to the time required to accelerate electrons to the TeV range ($~ 50 - 100$ years). The range $10-10^3$ MeV of the spectrum is overestimated by an extrapolation of our model, and we stop plotting the synchrotron spectrum beyond 10~MeV. We note that the standard synchrotron emissivity looses validity in this energy band as one approaches the radiation-reaction limit at $\approx 150\ \mathrm{MeV}$ \citep{Guilbert.etal:83}. A possible explanation is also that the acceleration processes operating in the Crab nebula can not reach electron energies higher than a PeV\footnote{PeV electrons radiate at $10\,\mathrm{MeV}$ in a $140$-$\mathrm{\mu G}$ field.}.
An additional feature not reproduced here is the re-brightening in the range $0.01 - 0.1$ eV (see Fig. \ref{F_TOT_tdep}) possibly due to dust contribution \citep{Meyer.etal:10}. 

\begin{figure}
	\includegraphics[width=0.5\textwidth]{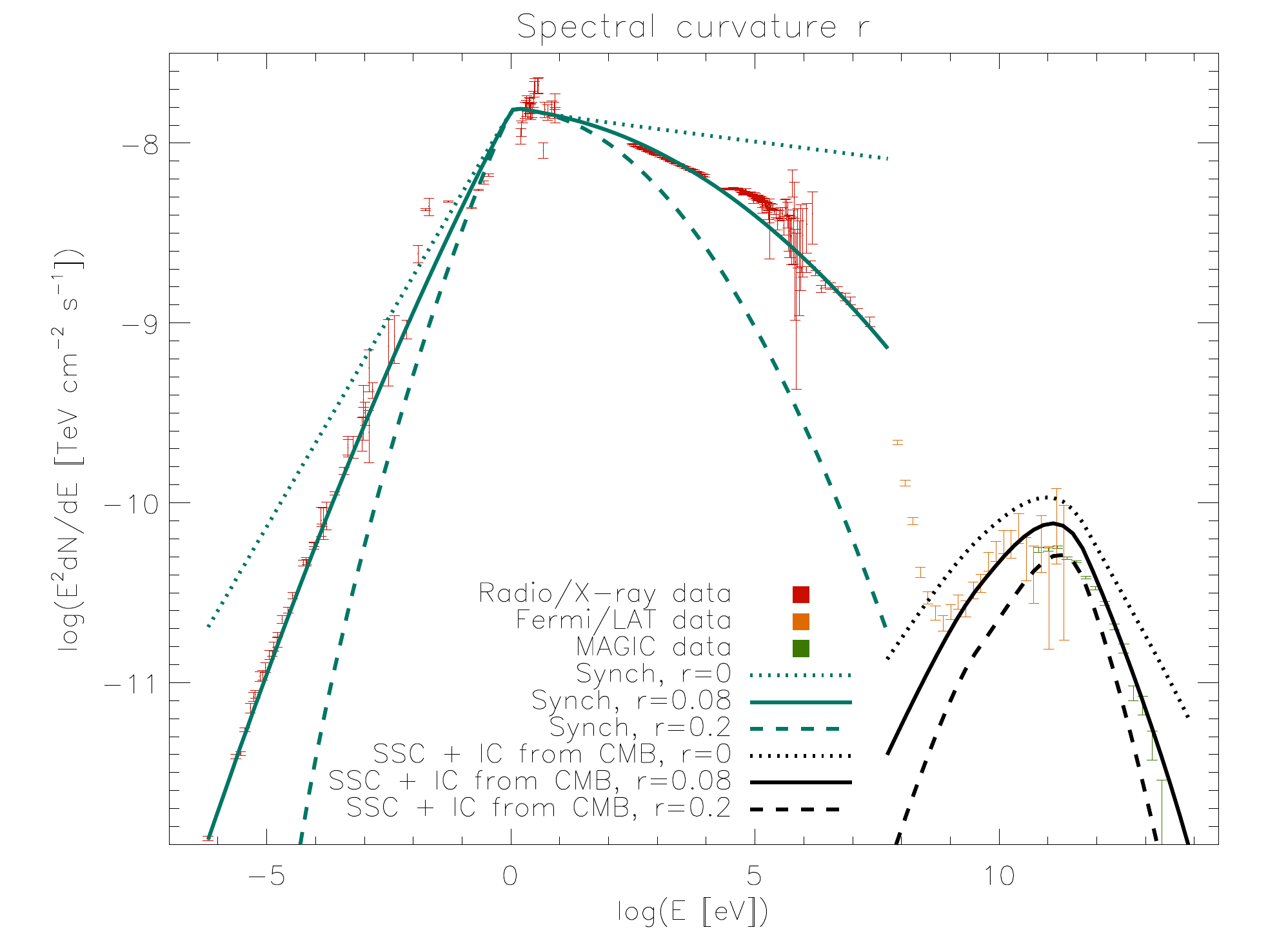}
\caption{Theoretical differential synchrotron photon spectra (Eq. \ref{f_syn_tdep}) and IC spectra (Eq. \ref{F_IC_TOT_tdep}) for three values of $r$ ($0.2, 0.08, 0$) are compared with the same radio and soft $X$-ray dataset as Fig.\ref{F_TOT_tdep}. For the other parameters the best-fit value in Fig.\ref{F_TOT_tdep} was used.}
\label{F_r_3}
\end{figure}

Figure \ref{F_r_3} shows the overall theoretical spectrum with three different values of $r$ ($0.2, 0.08, 0$). In particular, the $r=0$ curve, corresponding to a single power-law electron-injection spectrum (see Eq. \ref{N}), cannot reproduce the spectral curvature in the synchrotron range. To be noted from the figure is the mismatch of the $r=0$ curve both below and above the cooling break at $2.7$~eV, implying that at least two breaks in the injection spectrum, if not more, are required to reproduce the SED of the Crab nebula. As each break requires 2 free parameters, a combination of multiple broken power-laws energy spectra with ad-hoc energy breaks is an unlikely explanation for the synchrotron $\nu F_\nu$ spectrum, regardless of the consistency with the VHE spectrum. 
The main result shown by Fig. \ref{F_r_3} is that a single electron distribution accounts naturally for the curvature of the spectrum across several orders of observed photon energy with no need of energy breaks, provided one allows for log-parabola injection spectra. 

\begin{figure}
	\includegraphics[width=0.5\textwidth]{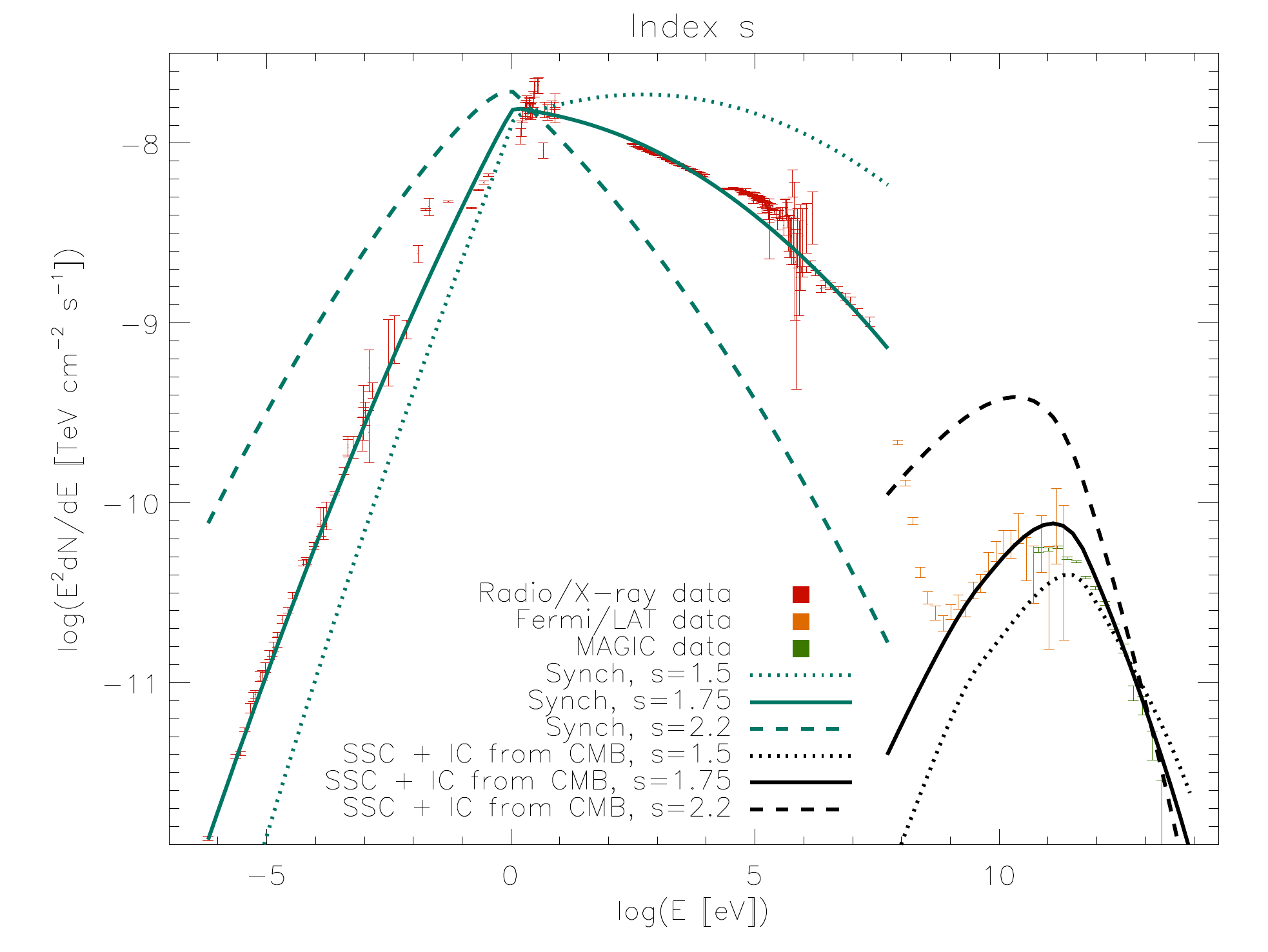}
\caption{Theoretical synchrotron photon spectra (Eq. \ref{f_syn_tdep}) and IC spectra (Eq. \ref{F_IC_TOT_tdep}) for three values of the index $s$ ($1.5$, $1.75$ and $2.2$) are compared with the same data sets as in Fig.\ref{F_TOT_tdep}. The parameters of the theoretical spectrum are the same as in Fig.\ref{F_TOT_tdep}. }
\label{F_s_3}
\end{figure}

\begin{figure}
	\includegraphics[width=0.5\textwidth]{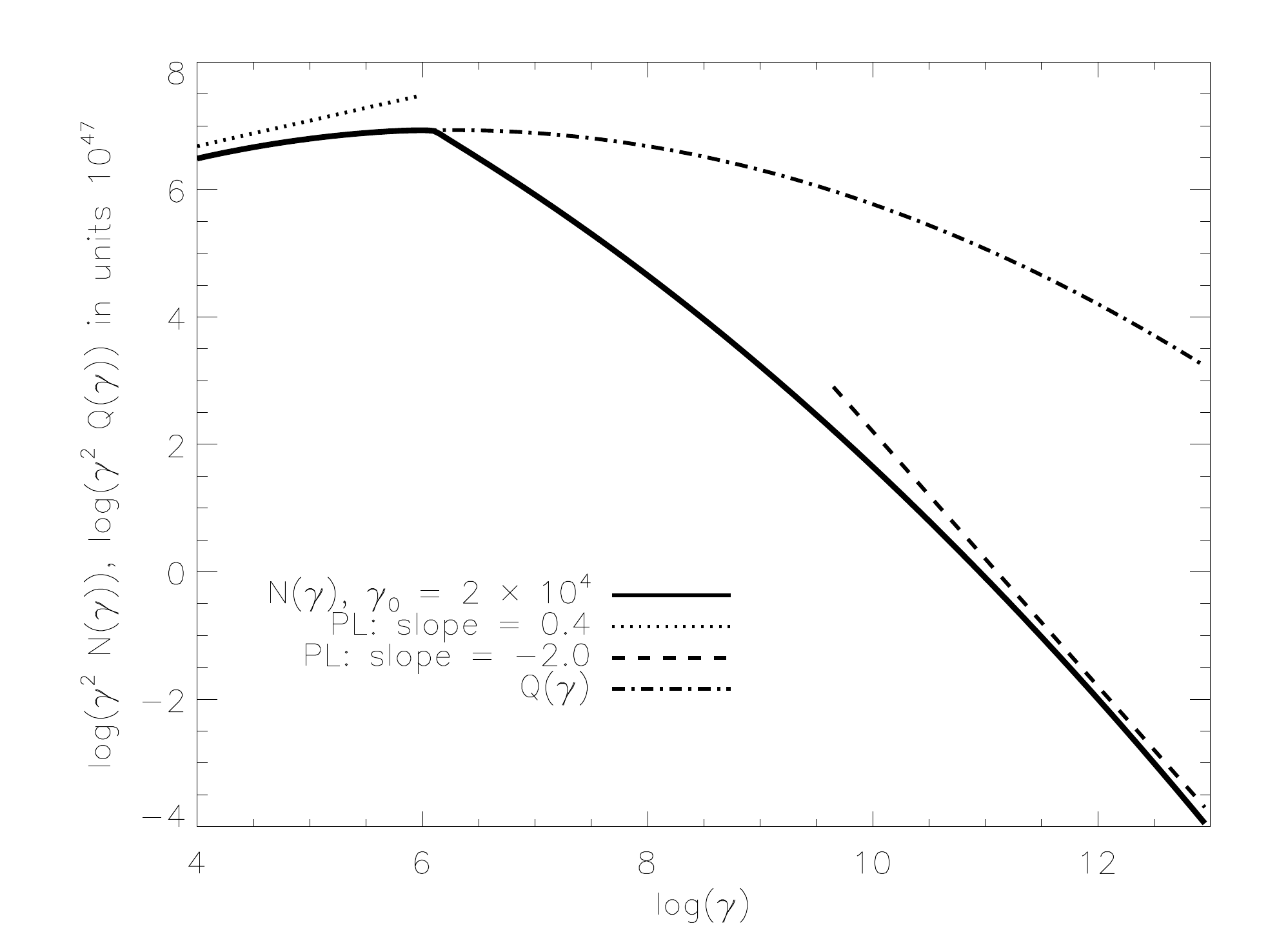}
\caption{Electron energy spectrum in Eq. \ref{espec} in units of $10^{47}$ multiplied by $\gamma^2$ (solid line) at $\gamma > \gamma_0$ compared with two asymptotic power-laws at low (slope $= 0.4$, dotted) and high (slope $= -2$, dashed) energy; here $s=1.75$, $r=0.08$. For comparison the differential electron production rate $Q(\gamma)$ defined in Eq. \ref{Q_gamma} is depicted.}
\label{dist_elec_2}
\end{figure}

Figure \ref{F_s_3} shows the steepening of the synchrotron photon spectrum as a result of the steepening of electron spectrum. The slope of the $\nu F_\nu $ spectrum in the radio range is expected to be the closest to that resulting from a single-power-law electron distribution as the logarithmic factor in Eq. \ref{f_syn_tdep} is very small ($\epsilon \simeq \epsilon_s \simeq 0.27 $ meV and $x \simeq 1$). Figure \ref{F_TOT_tdep} indicates a $\nu F_\nu $ spectral index of $\sim 0.64$ for the radio band, corresponding to an electron slope $\sim 1.72$, very close to the best-fit value of $s=1.75$.
In Fig. \ref{F_a_3} we explore the observational consequences of a possible inherent time-dependence of the particle injection by varying the parameter $a$ (see Eq. \ref{Q_a}) over the range $-0.5, -0.05, 0$. We can conclude that observations are consistent with a constant ($a=0$) or a slowly declining ($a=-0.05$) rate of particle injection. Figure \ref{dist_elec_2} compares the best-fit electron distribution $N(\gamma)$ from Eq. \ref{espec} with two asymptotic power laws. The overlaid differential electron production rate $Q(\gamma)$ from Eq. \ref{Q_gamma} shows the effect of the energy losses.

The parameter $\gamma_0$ is poorly constrained, because apart from a logarithmic factor any variation in $\gamma_0$ in Eq.~\ref{Q_gamma} can be compensated by adjustments of other parameters ($s$, $r$, or $N_0$). Technically, $\gamma_0$ may be interpreted the Lorentz factor of electron that return to the upstream region for the first time, and hence for highly relativistic shocks $\gamma_0$ can be very large. We note that the radio spectrum is best reproduced if we allow for curvature in the synchrotron spectrum down to $100\,\mathrm{MHz}$, corresponding to an electron Lorentz factor of about $10^3$. Our figures demonstrate that $\gamma_0=2\cdot 10^4$ is also an acceptable choice.  

\section{Discussion}\label{discussion}

The model of the broadband Crab nebula spectrum presented in the previous section relies on the {\it ansatz} of an energy dependence of the probability ${\cal P}$ that the multi-TeV energetic electrons undergoing acceleration at the wind TS can be confined in the proximity of the shock less efficiently as the particle energy grows. As mentioned above, such an assumption is also supported by the modelling of blazars $X$-ray spectra during the past decade, by the 3FGL catalog of extended astrophysical sources, and by {\it in-situ} measurements of interplanetary shocks. 

\begin{figure}
	\includegraphics[width=1.05\columnwidth]{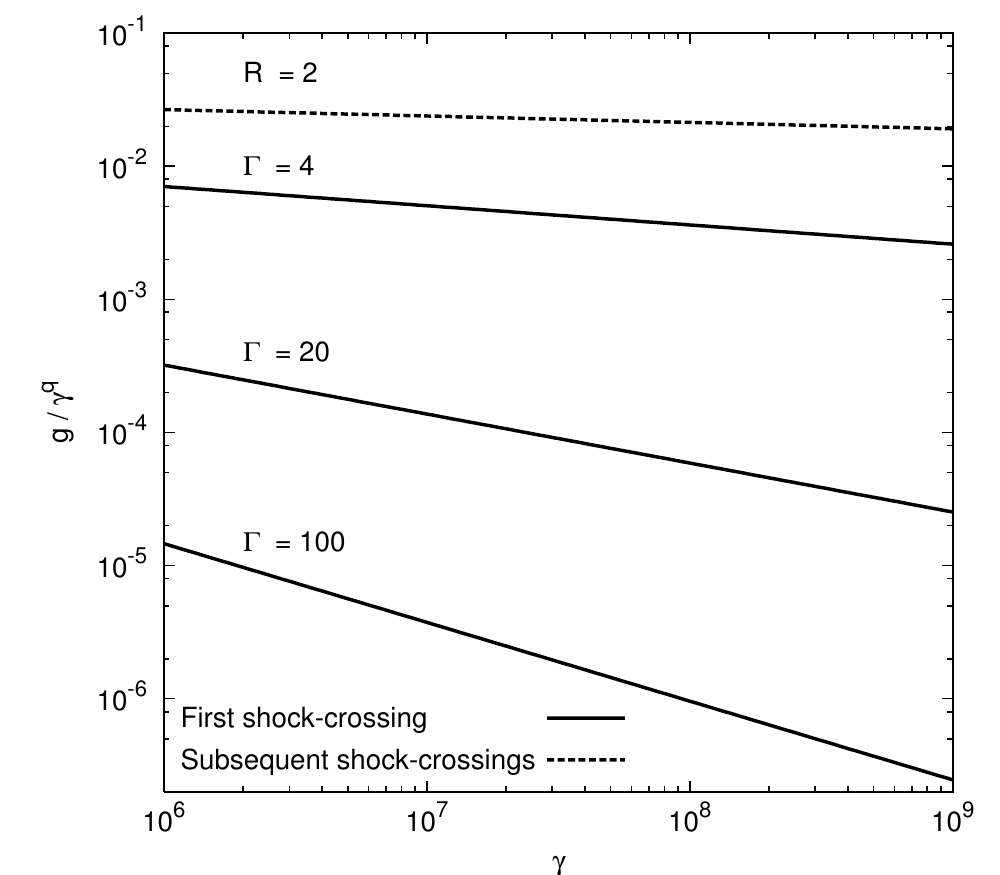}
\caption{Probability ${\cal P} (\gamma) = g \gamma ^{-q}$, as a function of $\gamma$, calculated using $q = 2 r {\rm log} (R)$ for three selected values of $\Gamma$ ($4$, $20$, $100$) for the first shock-crossing (solid lines), where $R \simeq \Gamma^2/2$, compared with ${\cal P} (\gamma)$ for subsequent shock-crossings ($R \simeq 2$, dashed line); here we have used the best-fit value $r = 0.08$.
\label{Prob}}
\end{figure}

By using a crude assumption, we can derive from the best-fit parameters of the observed photon spectrum an estimate of the microscopic parameter $q$. The electron-positron pair wind, possibly loaded by ions \citep{Hoshino.etal:92}, impinges on the TS with a speed customarily regarded as ultra-relativistic, although the order of magnitude of $\Gamma$ is substantially model-dependent and controversial. The value $\Gamma \simeq 10^6$ early proposed by \cite{Kennel.Coroniti:84}, with a single electron population injected at $1$ TeV, could not account for radio-to-IR observations. We adopt here $\Gamma \simeq 10^2$, consistent with a cold, magnetically dominated upstream wind, outward to the TS, that appears point-like in optical band \citep[for a discussion]{Kirk.etal:09}.  
For such a shock, the relation $r = q/2{\rm log} (R)$ yields $q \simeq 2 r {\rm log} (\Gamma^2/2)$ for the first shock crossing ($R \simeq \Gamma^2/2$) and $q \simeq 2 r {\rm log} (2)$ thereafter ($R \simeq 2$).  The assumed value $\Gamma \simeq 10^2$ leads to $q=0.59$. Such a small value of $q$ has nevertheless a dramatic impact on the dependence of ${\cal P}$ on the particle energy, indicating an extremely efficient depletion of the shock region. Figure \ref{Prob} depicts ${\cal P}(\gamma; \Gamma) = g/\gamma^{q(\Gamma)}$ where best-fit values are used for $g = \gamma_0^q \times 10^{-(s+(q-2)/2){\rm log} R}$. For completeness we show in Fig. \ref{Prob} also $g/\gamma^q$ for shock-crossings after the first one although the small values of ${\cal P}$ for first shock-crossing make subsequent crossings very unlikely to take place. For a non-relativistic strong shock (assuming a density compression equal to $4$) the relation $r = q/2{\rm log} (R) $ yields $q = 2 r U_1/v$, where we have used $R = 1+  \Delta p/ p $ and $U_1$ is the upstream fluid speed. Since the energy gain per first crossing for non-relativistic shocks is smaller than for ultra-relativistic shocks, the value of $q$ is likely to be typically smaller ($U_1/v \ll 1$), i.e. $q<0.1$.

In contrast with the infinitely planar shock case, the probability ${\cal P}$ drops dramatically with the flow $\Gamma$. Monte Carlo test-particle simulations for infinitely planar shocks in \cite{Lemoine.Pelletier:03} found a negligible dependence of the probability of return, $P(v)$, on $\Gamma$ and a mild dependence on $\gamma$. By introducing the {\it ansatz} ${\cal P} = g/\gamma^q$, we argue that at VHE the non-planar large-scale structure of the shock likely affects the capability of the shock of containing the most energetic particles.

In terms of macroscopic interpretation, we consider the case that electrons undergo scattering in the fluid upstream of the shock \citep{Achterberg.etal:01,Pelletier.etal:09} so that the relevant physical length-scale is the scale of confinement of particles upstream or the diffusion-advection scale, defined as $D/c$, where $D$ is the transport coefficient in the average direction of the flow impinging on the shock, that is a radial outflow at the TS to large-scale; $D/c$ has to be larger than the scattering mean free path in order to guarantee particle confinement, hence acceleration. In the downstream fluid the turbulence has to be strong enough to enable the return of the electrons back to the shock. At the wind TS, the large-scale magnetic field is quasi-perpendicular to the flow direction\footnote{At non-relativistic quasi-perpendicular shocks, the acceleration is very fast and efficient \citep{Jokipii:87} as confirmed by test-particle simulations at fluid-scale \citep[and reference therein]{Fraschetti.Giacalone:15} and by hybrid simulations at ion kinetic scale \citep{Giacalone:05} for interplanetary shocks travelling into pre-existing turbulence. Such result is also confirmed via hybrid simulations for high-Mach number supernova remnant shocks travelling, e.g., in partially ionized fluid \citep{Ohira:16}.}; thus, $D$ is well approximated by $D_\perp$ at large-scale. 
In a quasi-perpendicular configuration we do not expect self-generated waves ahead of the ultra-relativistic shock to enhance the scattering of particles and thus keeping them close to the shock, although streaming of particles could be relevant in regions of the TS with a quasi-parallel magnetic field \citep{Vainio.etal:04}; thus, $D_\perp$ is mainly due to pre-existing wind turbulence. 
We can compare $D_\perp/c$ with the typical TS scales. We adopt for the sake of simplicity the Bohm scaling, i.e., namely scattering mean free path equals $r_g$ determined by the unperturbed magnetic field. The Bohm scaling generally applies to weak field regime where the power of the magnetic fluctuations along all three space directions is nearly isotropic and corresponds to large-scale diffusion coefficients comparable in the direction parallel and perpendicular to the field. In the Bohm regime $D/c = r_g/3$; thus, the scattering mean free path ($\sim r_g$) becomes comparable with $D/c$, the scattering becomes inefficient, and particles are likely to leave the system. The Bohm assumption sets an upper limit on $D_\perp$: if the turbulence is weaker \citep[e.g.]{Fraschetti.Giacalone:12}, {$D_\perp$} will be smaller and  $ D_\parallel$ will be larger respectively leading to smaller values of the diffusion-advection scale and larger values of the the scattering mean free path, thereby enhancing the particle leak-out. 

We have shown that the synchrotron spectrum is produced by electrons in the GeV - TeV range 
orbiting around a large-scale magnetic field with strength $B_0  = 140 \, \mu$G. The value of $B_0$ probes the region behind and close to the shock where cooling length-scale of TeV electrons is small; the field $B_0$ might be the result of local amplification due to the inhomogeneities of the pulsar wind advected through the shock (see \cite{Mizuno.etal:14} for relativistic and \cite{Fraschetti:13} for non-relativistic flows) and the average nebular field might be smaller than $B_0$. 
An electron emerging upstream will have an energy in the upstream frame $E' = \Gamma E /\sqrt{2}$, where $\Gamma/\sqrt{2}$ is the Lorentz factor of the downstream fluid relative to the upstream frame \citep{Blandford.McKee:76}; if the magnetic field $B_0$ is only the result of shock compression from the jump conditions at a quasi-perpendicular shock front, the upstream field in the upstream frame is $B' \sim B_0/{\sqrt{r^2-1}\Gamma}$, where the ``$\sim$'' accounts for the local change of magnetic obliquity along the shock surface and $r = 3$ is the density compression for high Mach number ultra-relativistic shocks. Thus, by taking $B_0  = 140 \, \mu$G, the upstream electron gyroradius in the upstream frame is given by $r'_g = E' / e B' \simeq 2 \sqrt{2} \Gamma^2 E/e B_0 \sqrt{2} \simeq 1.5 \times 10^{-1}$ pc for $E \simeq 1$ TeV, where $e$ is the electron charge and $\Gamma \simeq 10^2$ (shrinking to $ r'_g \simeq 1.5 \times 10^{-3}$ pc at $E = 10$ GeV). The resulting diffusion-advection length in the upstream wind frame at $1$ TeV is $D/c = r_g'/3 = 5 \times 10^{-2}$ pc. Particles in the cold wind upstream are confined within a region much smaller than the radius of the nebula TS ($\sim 0.1$ pc), typically identified with an equatorial $X$-ray ring comprised of a collection of knots \citep{Hester.etal:02}.

\section{Conclusions}\label{conclusions}

We have developed a simple analytic one-zone model for the broadband baseline photon spectrum of the Crab nebula. We find an impressive agreement with observation over the entire range $(10^{-5} - 10^{14})$ eV, except the MeV-range and an IR re-brightening at $\sim 0.01-0.1$ eV. The energetic electrons, injected at $\sim 10$ GeV, are assumed to be accelerated to a log-parabola distribution at the nebula wind termination shock. The electron spectrum is then obtained as solution to the time-dependent continuity equation accounting for energy losses and a possibly time-dependent injection rate. This model can account for the broad and flat IC peak matching the Fermi/LAT emission beyond $1$ GeV with the VHE emission up to tens of TeV and the synchrotron emission in the radio up to soft $X$-ray range. We 
show that the resulting synchrotron emission accounts naturally for the curvature of the broadband observed photon spectrum. Our broadband model is consistent with a uniform magnetic field throughout the nebula $B_0 = 140 \, \mu$G, in agreement with previous MHD models \citep{Meyer.etal:10}, although this field strength can also result from local downstream amplification.  
Simple assumptions on the large-scale perpendicular topology of the magnetic field at the nebula termination shock allow to infer the energy dependence of the probability that the energetic electrons remain in the proximity of the shock. The model hereby proposed is applicable to a broad variety of sources (pulsar wind nebulae, supernova remnants) over an unprecedentedly broad energy range.

\section*{Acknowledgements}

The work of FF was supported, in part, by NASA under Grants NNX13AG10G, NNX15AJ71G. We thank the referee for useful comments. We thank D. Mazin, N. Otte and M. Holler for sharing low-energy, VERITAS and HESS II data respectively. FF acknowledges J. R. Jokipii, J. Giacalone and J. K\'ota for continuing discussions and E. J. Summerlin for email exchange. FF thanks for the hospitality the Harvard/Smithsonian Center for Astrophysics where part of this work was performed.





\def \apss{{\it Astrophys.\ Sp.\ Sci.}}
\def \aj{{\it AJ}}
\def \apj{{\it ApJ}}
\def \apjl{{\it ApJL}}
\def \apjs{{\it ApJS}}
\def \araa{{\it Ann. Rev. A \& A}}
\def \prc{{\it Phys.\ Rev.\ C}}
\def \aap{{\it A\&A}}
\def \aaps{{\it A\&ASS}}
\def \mnras{{\it MNRAS}}
\def \physscr{{\it Phys.\ Scripta}}
\def \pasp{{\it Publ.\ Astron.\ Soc.\ Pac.}}
\def \gca{{\it Geochim. Cosmochim.\ Act.}}
\def \nat{{\it Nature}}
\def \solphys{{\it Sol.\ Phys.}}


\bibliographystyle{mnras}
\bibliography{ms}




\appendix


\bsp	
\label{lastpage}
\end{document}